\documentclass[aps,prl,preprint,superscriptaddress,twocolumn,10pt,fleqn]{revtex4-1}

\newcommand{\bra}[1]{\left\langle #1\right|}
\newcommand{\ket}[1]{\left| #1\right\rangle}
\newcommand{\be}[0]{\begin{equation}}
\newcommand{\ee}[0]{\end{equation}}
\newcommand{\tr}[0]{{\rm Tr}}
\newcommand{\re}[0]{{\rm Re}}
\newcommand{\im}[0]{{\rm Im}}
\newcommand{\bml}[0]{\begin{multline}}
\newcommand{\eml}[0]{\end{multline}}
\newcommand{\nn}[0]{\nonumber}
\newcommand{\dom}{\delta \omega}
\newcommand{\bea}{\begin{eqnarray}}
\newcommand{\eea}{\end{eqnarray}}

\usepackage[fleqn]{amsmath}
\usepackage{graphicx}
\usepackage{float}
\begin{document}
\title{Complete temporal characterization of a single photon}

\author{Zhongzhong Qin}
\affiliation{Institute for Quantum Science and Technology, University of Calgary, Alberta T2N1N4, Canada}
\affiliation{Quantum Institute for Light and Atoms, State Key Laboratory of Precision Spectroscopy, East China Normal University, Shanghai 200062, People's Republic of China}
\author{Adarsh S. Prasad}
\affiliation{Institute for Quantum Science and Technology, University of Calgary, Alberta T2N1N4, Canada}
\author{Travis Brannan}
\affiliation{Institute for Quantum Science and Technology, University of Calgary, Alberta T2N1N4, Canada}
\author{Andrew MacRae}
\affiliation{Department of Physics, University of California, Berkeley, California 94720, USA}
\author{A. Lezama}
\affiliation{Instituto de F\'{i}sica, Facultad de Ingenier\'{i}a, Universidad de la Rep\'{u}blica, J. Herrera y Reissig 565, Montevideo 11300, Uruguay}
\author{A. I. Lvovsky}
\email{LVOV@ucalgary.ca}
\affiliation{Institute for Quantum Science and Technology, University of Calgary, Alberta T2N1N4, Canada}
\affiliation{Russian Quantum Centre, 100 Novaya St., Skolkovo, Moscow 143025, Moscow, Russia}
\date{\today}
\begin{abstract}
Precise information about the temporal mode of  optical states is crucial for optimizing their interaction efficiency between themselves and/or with matter in various quantum communication devices. Here we propose and experimentally demonstrate a method of determining both the real and imaginary components of a single photon's temporal density matrix by measuring the autocorrelation function of the photocurrent from a balanced homodyne detector at multiple local oscillator frequencies. We test our method on single photons heralded from biphotons generated via four-wave mixing in an atomic vapor and obtain excellent agreement with theoretical predictions for several settings.\\
%\vspace{5 ex}
\end{abstract}

\pacs{Valid PACS appear here}% PACS, the Physics and Astronomy

\maketitle

\noindent\textbf{Introduction}\\
Single photons and single photon qubits are among the foundations of most quantum optical information processing techniques such as cryptography \cite{Gisin2002crypto}, teleportation \cite{QTexp}, repeaters \cite{DLCZ} and computing \cite{KLM}. Many of these applications require the photons to have a well-defined, pure modal structure. Possessing precise information about that structure is essential for quantum-optical technology.

An approximate guess of a photon's mode can be inferred theoretically from the characteristics of the source \cite{grosshans2001effective,aichele2002optical,Mat04,Molmer06,SasakiSuzuki06}, but this information is not always available or reliable. For example, this approach would not work for photons sent in by a remote party in a communication scheme, or for photons from an incompletely characterized mesoscopic source. Therefore, it is important to have a technique for precise characterization of a photon's mode experimentally. While such techniques are relatively well developed for spatial modes \cite{Lvo09Rev,lundeen2011}, their extension into the temporal domain is challenging.

One approach to studying the temporal structure of the photon would be to look at the photon detection event statistics as a function of time. For example, this approach has been used to study the timing of coherent double Raman scattering from an atomic ensemble \cite{Kimble04}. Further insight into the photon preparation quality can be gained by studying time-dependent photon counting autocorrelation statistics \cite{Solomon12}. However, these techniques provide no information about the phase coherence between different segments of the photon's temporal mode.

Complete information about a photon's temporal properties can be obtained by studying its interference with a classical field. Polycarpou {\it et al.} used adaptive waveform shaping of local oscillator (LO) pulses \cite{Bel12} to heuristically find the LO temporal mode that maximizes the efficiency of homodyne detection of the photon. This occurs when the LO temporal mode matches that of the signal, enabling measurement of that mode. However, physical shaping of LO pulses is quite sophisticated experimentally. Furthermore, this technique has only been demonstrated for pure temporal modes.

An alternative approach to measuring the spectral density matrix of the photon has been proposed in Ref. \cite{Wojciech}. It is based on bringing the photon into interference with a pair of weak coherent pulses with varied separation between them. However, this method can only be applied to ultrashort pulses whose width is a few optical cycles. Furthermore, the experimental test in Ref. \cite{Wojciech} has been performed on a thermal state rather than the single photon state.

For a photon with bandwidth resolvable by detection electronics, the time domain statistics may be measured directly and in real time by analyzing the time-dependent statistics of the homodyne detector's output photocurrent with a continuous-wave LO. MacRae {\it et al.} \cite{And12thesis,And12PRL} showed that the autocorrelation function of this photocurrent estimates the real part of the density matrix defining the photon's temporal mode. Subsequently, this approach has been utilized for the ``Schr\"{o}dinger cat" and two-photon Fock states \cite{Mor13}.

However, this method does not yield any information about the imaginary part of the photon's temporal density matrix (TDM). In this paper we present an experimental technique of \emph{polychromatic optical heterodyne tomography}, which relies on acquiring the autocorrelation data of the homodyne photocurrent at multiple LO frequencies. The method enables us to determine both the real and imaginary parts of the photon's TDM, or, equivalently, both its amplitude and phase, thereby completely characterizing its temporal state. It works equally well for pure and mixed temporal modes. %Recently, it has also been shown that two-mode Gaussian states showing strong energy imbalance between spectral sideband modes cannot be completely reconstructed by spectral homodyne detection but can be inherently revealed by resonator detection \cite{Vil13PRL,Vil13PRA}.

%Before we proceed to describing the method, we note that the task pursued here is complementary to that of homodyne tomography \cite{Lvo09Rev}. In the latter case, the goal is determining the state of the electromagnetic field in a given mode, whereas our goal here is determining the mode of a given state (single photon). Interestingly, both task are addressed using similar experimental tools. %Furthermore, in the case of multiple spectral modes,

\vspace{2 ex}
\noindent\textbf{Materials and Methods}\\
The (pure) temporal mode of a photon is defined by annihilation operator
\be
\label{definition}
\hat{A}_\phi = \int_{-\infty}^{\infty} \hat{a}_{t} \phi(t) {\rm d}t,
\ee
where $\phi(t)$ is  the temporal mode function (TMF) and $\hat{a}_{t}$ represents the instantaneous annihilation operator at time $t$. Although single photons associated with a certain moment in time are ill-defined, treatment \eqref{definition} is approximately valid as long as the spectral width of the photon is much less than its frequency \cite{Fedorov}. A single photon state in this mode is then given by $\ket{1_\phi} = \hat{A}_\phi^\dag \ket{0} = \int_{-\infty}^{\infty} \phi^*(t)\ket{1_t} {\rm d}t$, where $\ket{1_t}=\hat a_t^\dag \ket{0}$.

%\footnote{Although single photons associated with a certain moment in time are ill-defined, treatment \eqref{definition} is approximately valid as long as the spectral width of the photon is much less than its frequency. See M.V. Fedorov {\it et al.}, Phys. Rev. A {\bf 72}, 032110 (2005) for a detailed discussion of this matter.}

The digital nature of the data acquisition system used in our experiment compels us to represent the temporal modes in terms of discrete time bins. The single photon state in temporal mode $\phi(t)$ can then be approximately expressed as $\ket{1_\phi} = \sum_j \phi^*(t_j) \ket{1_j}$ with $\sum_j \vert\phi(t_j)\vert^{2}=1$. Here $t_j$ is the time associated with the $j^{\rm th}$ bin and  $\ket{1_j}$ is the state containing one photon in the top-hat temporal mode associated with the $j^{\rm th}$ bin and vacuum in all other bins. The density operator of the photon is then represented as $\sum_{mn} \rho_{mn} \ket{1_m} \bra{1_n}$ where $\rho_{mn}$ is the temporal density matrix.

%The creation operator for the $k^{th}$ time bin is $\hat{a}_k^\dag$ verifying: $\hat{a}_k^\dag \ket{n_j}= (1-\delta_{jk})\ket{n_j} + \delta_{jk} \sqrt{n+1} \ket{(n+1)_j}$.

The homodyne current for the $j^{\rm th}$ time bin $I(t_j)$ is proportional to the quadrature \be\label{quaddef}
\hat{X}_j = (\hat a_j e^{-i\theta_j} +\hat a_j^\dag e^{i\theta_j})/\sqrt{2},
\ee
where $\theta_j = \dom \cdot t_j+ \theta_0$ is the optical phase difference between the LO and the signal. Here $\dom$ is the frequency detuning between the LO and the signal and $\theta_0$ the LO relative phase at $t=0$. The autocorrelation matrix for the homodyne current is then
\bea
\label{Ajk}
\langle I(t_j)I(t_k) \rangle &\propto & \langle  \hat{X}_j \hat{X}_k \rangle = \tr[\hat{\rho}\hat{X}_j \hat X_k] \nn \\
&=& \sum_{mn} \rho_{mn} \bra{1_n}\hat{X}_j \hat X_k \ket{1_m},
\eea
where each matrix element can be evaluated using Eq.~(\ref{quaddef}) as
\begin{align}
\label{innerxjxk}
&\hspace{-1cm}\bra{1_n}\hat{X}_j \hat X_k \ket{1_m}  \\
&\hspace{-1cm}= \frac{1}{2} \Bigg[e^{-i\dom (t_k-t_j)}\delta_{km} \delta_{nj}   + \delta_{jk}\delta_{nm}+ e^{-i\dom (t_j-t_k)} \delta_{jm}\delta_{nk} \Bigg].\nn
\end{align}
Note that Eq.~(\ref{innerxjxk}) does not depend on $\theta_0$ due to the phase uncertainty of Fock states.

From Eqs.~\eqref{Ajk} and \eqref{innerxjxk}, one can obtain
\begin{equation}
\langle  \hat{X}_j \hat{X}_k \rangle  = \frac{1}{2} \delta_{jk} + A_{jk}.\label{autocorr}
\end{equation}
The first term in Eq.~\eqref{autocorr} corresponds to the autocorrelation matrix for the vacuum. The second term, which we call the \emph{reduced} autocorrelation matrix, is directly related to the photon's TDM:
\begin{equation}
A_{jk}= \re[\rho_{jk}] \cos[\dom (t_j-t_k)] + \im[\rho_{jk}] \sin[\dom (t_j-t_k)]. \label{autocorr-theo}
\end{equation}
If the LO frequency is same as that of the signal, i.e., at $\dom = 0$, the autocorrelation matrix depends only on the real part of the TDM. However, by using $\dom \neq 0$ one obtains access to its imaginary part.

In a realistic experiment, the photon being tested may experience losses, resulting in admixture of the vacuum into the  state detected. Our technique would still apply to this case, but the second term would enter Eq.~(\ref{autocorr}) with the coefficient equal to the transmissivity of the lossy element. For high losses, acquisition of larger quantities of data may be necessary to reduce the statistical uncertainties (detailed further in the Supplementary Material).

\begin{figure} [!bt]
\includegraphics[width=\columnwidth]{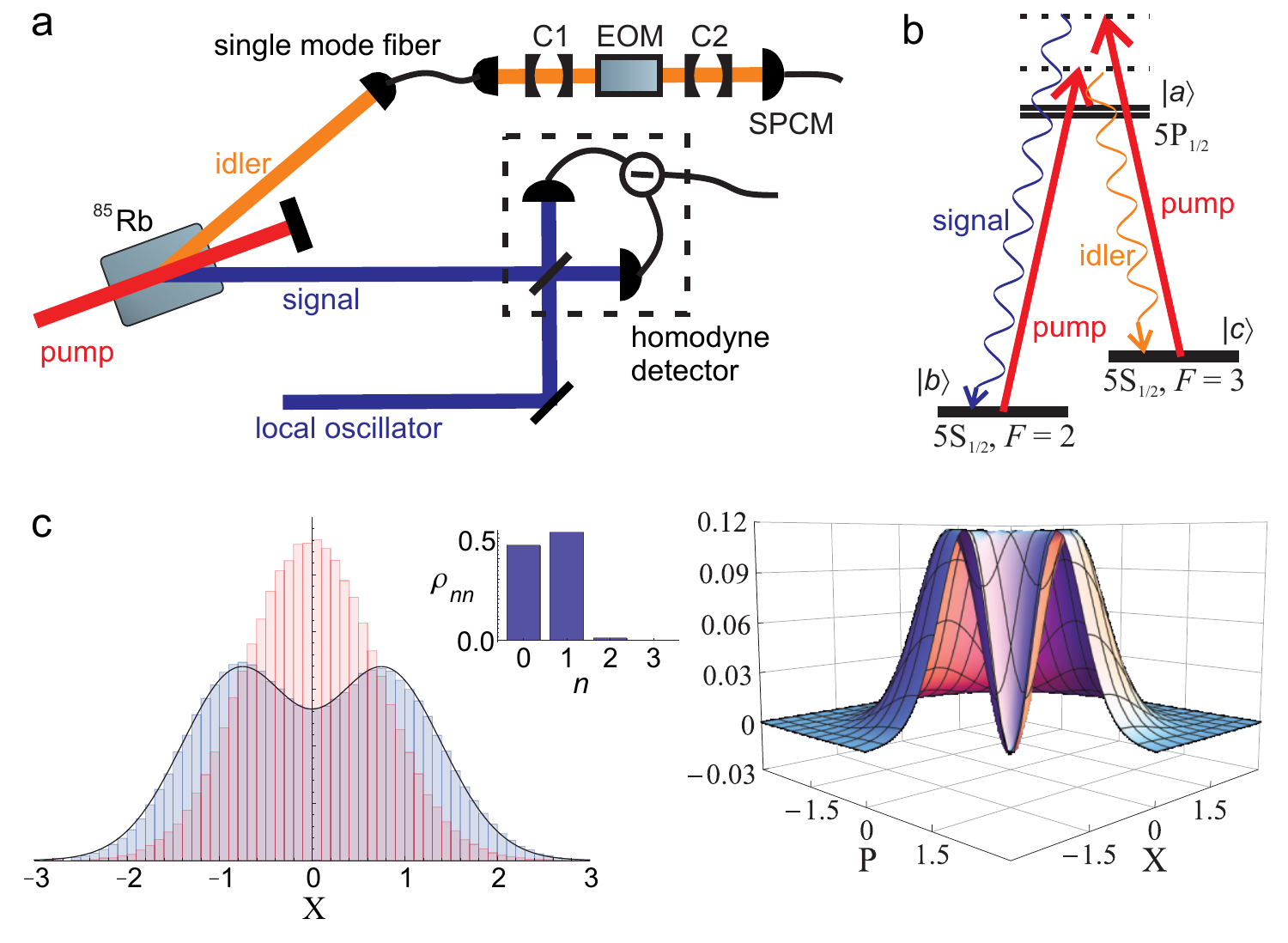}
\caption{\label{exptsetup}(a) Schematic of the experimental set-up (C1, C2: filter cavities; EOM: electro-optic modulator; SPCM: single photon counting module). The signal (blue) goes to the homodyne detector, whereas the idler (orange) passes through C1 (55 MHz) and C2 (7 MHz) before detection via the SPCM. The EOM between  C1 and C2 is optional. (b) The ${}^{85}$Rb three-level $\Lambda$ system, with the fields' configuration shown. (c) Reconstruction of the state of the electromagnetic field in the temporal mode determined experimentally (unmodulated case). Left to right: experimental quadrature distribution (blue) overlaid with that for the vacuum state (red); diagonal elements of the Fock-basis density matrix; Wigner function. The single-photon fraction is 52.9\%.}
\end{figure}

Our experimental scheme for creating the single-photon state and measuring the autocorrelation matrix is shown in Fig.~\ref{exptsetup}. We use coherent double Raman scattering (four-wave mixing) in an ensemble of $\Lambda$-type atoms to generate a two-mode squeezed state in a non-degenerate phase-matched configuration \cite{And12PRL}. A hot ${}^{85}$Rb vapor cell is pumped by a 1 Watt laser beam at 795 nm derived from a continuous-wave Ti$:$Sapphire laser. The signal and idler beams are spatially separated from the pump, and a specific spatial mode is selected in the idler channel using a single-mode fiber. Subsequently, the idler channel is subjected to spectral filtering by means of a lens cavity (C1) of a 55 MHz bandwidth \cite{Pan12} and a conventional Fabry-Perot cavity  (C2) of bandwidth $\gamma/2\pi= 7$ MHz. The usage of two cavities with incommensurate free spectral ranges ensures that the combined spectral filter has a single transmission peak of 7 MHz width. This results in a heralded photon with a temporal mode that can be easily resolved by our homodyne detector with a 100 MHz bandwidth \cite{Ran12}.

The idler beam is then coupled to a PerkinElmer single photon counting module (SPCM) with a dark count rate below 100 Hz. Both cavities are maintained at a stable frequency by using an alignment beam which is unblocked every few seconds to monitor and readjust the cavity resonance frequency. Detection of an idler photon projects the signal onto a single photon in a well-defined spatio-temporal mode conjugate to the idler. This signal channel is mode-matched with a continuous-wave LO (18 mW) for homodyne detection  \cite{Ran12}. The LO is derived from a diode laser that is locked and phase stabilized with respect to the pump using an optical phase-lock loop \cite{App09}.

\begin{figure}[t]
\includegraphics[width=\columnwidth]{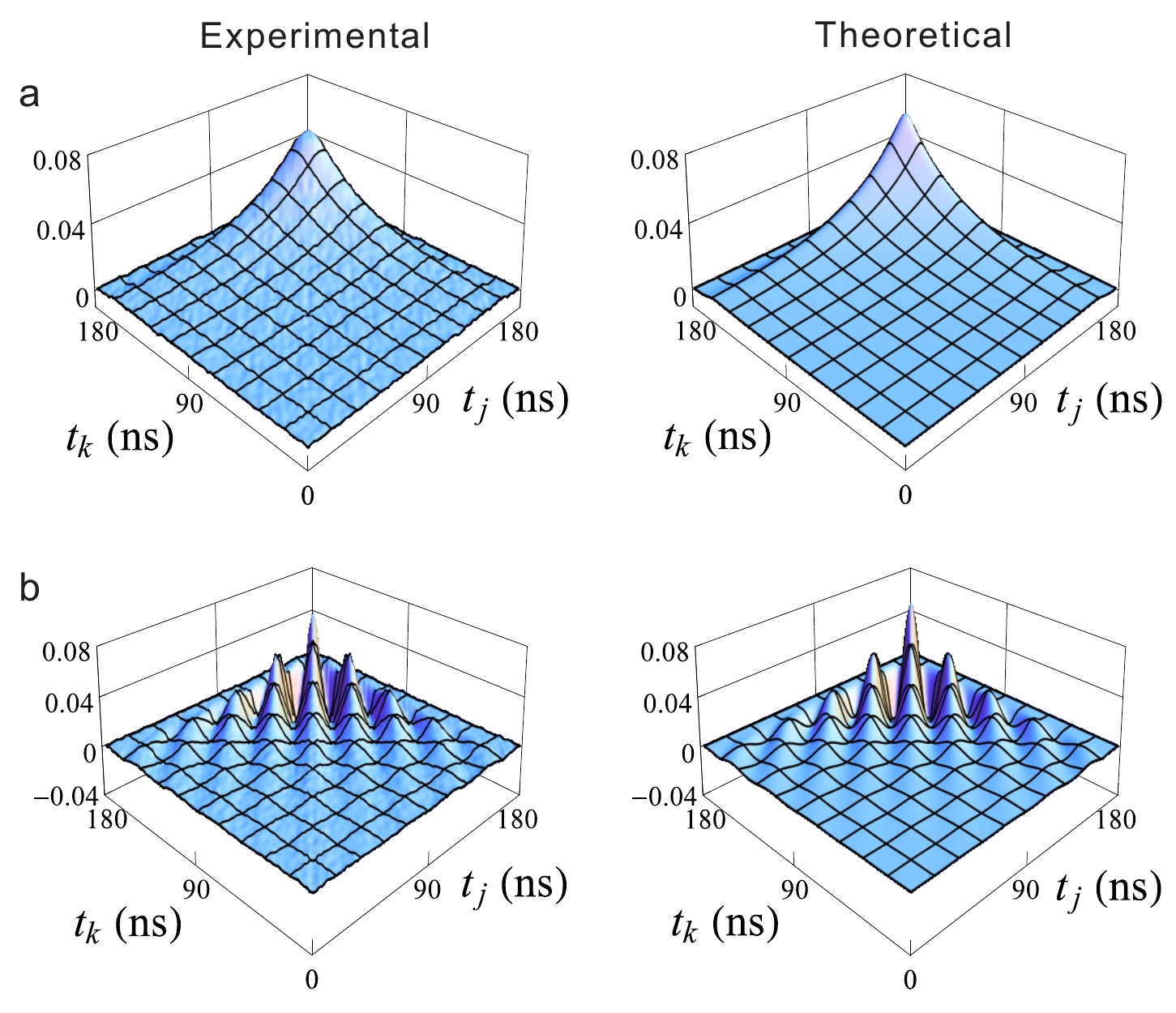}
\caption{\label{Autocorr-fig}Theoretical (right) and experimental (left) reduced autocorrelation matrices, for two different LO detunings:  (a) 0 MHz and (b) 27 MHz, corresponding to the measurement setting without modulation. The trigger photon arrives at $t=155$ ns.}
\end{figure}

A click from the SPCM in the idler channel acts as the trigger for the measurement of the signal. The homodyne photocurrent is recorded for 360 ns around the trigger point as reference with a time binning of 2 ns.  For each LO detuning, the autocorrelation matrix (\ref{Ajk}) of the homodyne photocurrent is obtained by taking an average over 2 million traces.
%We also record the same number of traces for the vacuum state and for randomly triggered thermal background.

Theoretically, the data corresponding to two LO detunings would constitute a quorum for the temporal mode reconstruction. Experimentally, however, we take data at eight different detunings to avoid the situation where the sinusoids in Eq.~\eqref{autocorr-theo} approach zero for all detunings simultaneously and to enhance the statistical accuracy of the recovered $\hat{\rho}$.

Once the autocorrelation matrices have been acquired, we process them to eliminate the vacuum term in Eq.~\eqref{autocorr}, as well as any contributions from the DC bias in the homodyne photocurrent and thermal background. These contributions are not correlated with trigger events, and are only dependent on the difference $t_j-t_k$. They can therefore be evaluated as the mean autocorrelation value along lines $t_j-t_k = \text{const}$ for the data points acquired significantly after the trigger pulse where no signal photon is expected. Subtracting them from the autocorrelation matrix yields the reduced autocorrelation matrix (Fig.~\ref{Autocorr-fig}) (detailed further in the Supplementary Material). %The theoretical autocorrelation matrices are determined by first calculating the theoretical TDM from Eqs.~(\ref{TMF}-\ref{TDM}) and then using the expression given by Eq.~\eqref{autocorr-theo}.

The TDM can now be determined by solving Eq.~\eqref{autocorr-theo} for each pair  $(j,k)$. However, such direct approach does not ensure positivity and normalization of the reconstructed density operator. To incorporate these \emph{a priori} constraints into the reconstruction, we implement a more sophisticated iterative optimization algorithm. The algorithm uses the eight experimental reduced autocorrelation matrices as the training set.  The difference between the experimental left-hand side of Eq.~\eqref{autocorr-theo} and the right-hand side of that equation evaluated from the estimated TDM, squared and summed over all pairs $(j,k)$ and all LO frequencies, is used as the cost function. Iterations utilize the diagonal representation of the TDM: $\hat\rho=\sum_i p_i\ket{\psi_i}\bra{\psi_i}$. In the first step of each iteration, the eigenvalues $p_i$ are adjusted to minimize the cost function while keeping them real, non-negative and totalling 1. In the second step, the eigenvectors $\ket{\psi_i}$ are optimized by pairwise unitary transformations. The process is repeated until the cost function asymptotically converges to give the best fit of the TDM.

\begin{figure*}[htbp]
\includegraphics[width=\textwidth]{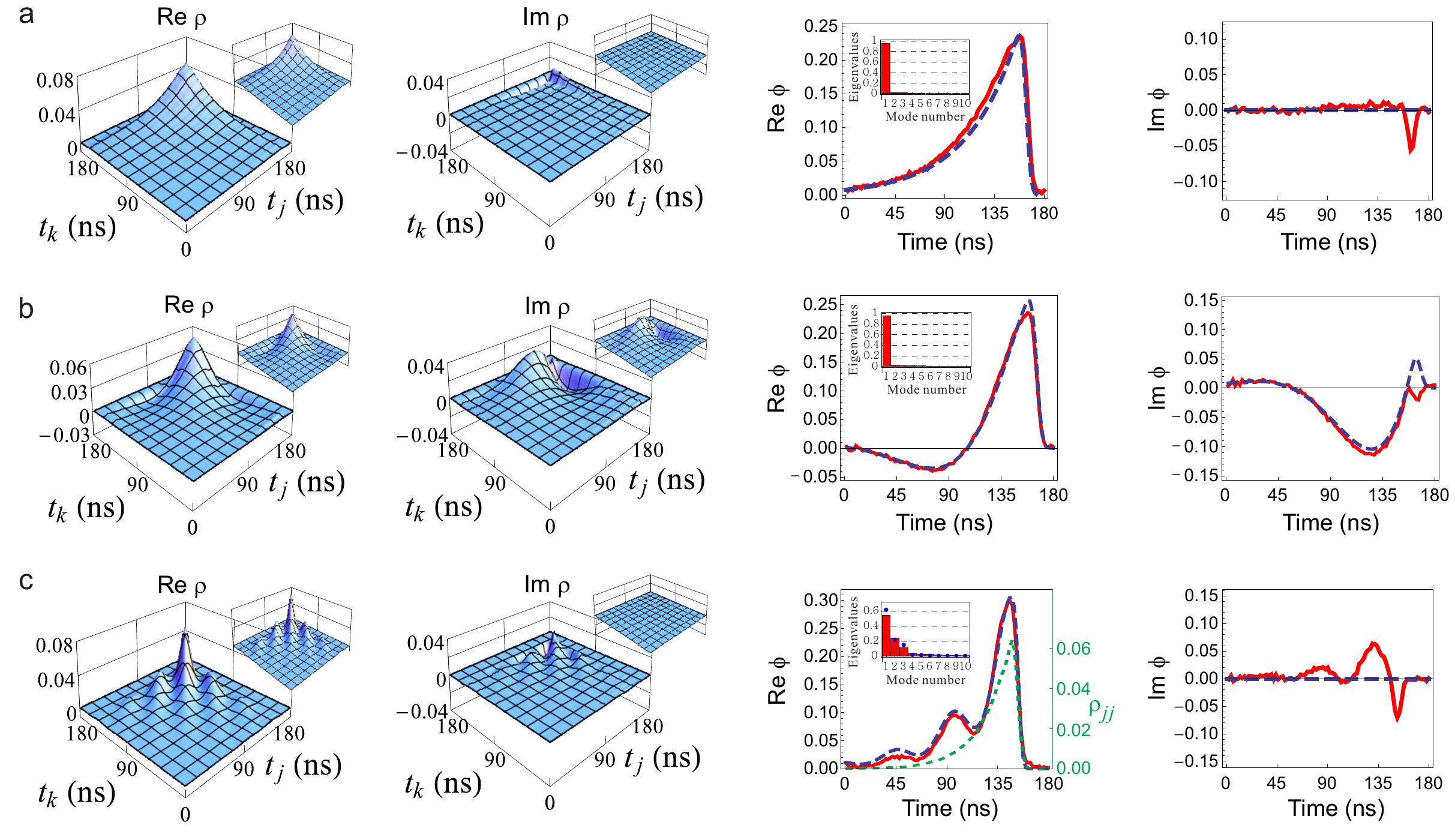}
\caption{\label{plots}Experimentally reconstructed temporal modes and their theoretical predictions for the cases without modulation (a), with virtual phase modulation (b), and with phase-randomized EOM modulation (c). For each case, the first and second panels show the real (first) and imaginary (second) parts of the TDM, with the insets showing corresponding theoretical plots. The third and fourth panels show the TDM's primary eigenvector as reconstructed from experimental data (solid red) and theoretical modeling (dashed blue). The green dotted line in (c) also shows the diagonal of the TDM, which is not affected by the modulation as expected from Eq.~(\ref{TDMEOM}). The insets in the third panels show the distribution of eigenvalues (red bars) obtained experimentally. Without EOM modulation (a, b), the theoretically expected mode is pure so the TDM is expected to have only one nonvanishing eigenvector. In the case with EOM modulation (c), the TDM is mixed and the solid blue dots in the inset show theoretical eigenvalues. The trigger photon arrives at $t=155$ ns for (a), (b) and at $t=145$ ns for (c).}
\end{figure*}

The theoretically expected mode is calculated from the properties of our experimental setup. The primary element determining the mode of the heralded photon is the narrowband filter cavity C2 in the idler channel. Additionally, the mode's bandwidth is limited by the $\sim 50$ MHz gain bandwidth of the four-wave mixing process used to generate the biphotons. This effect is taken into account in theoretical plots in Figs.~\ref{Autocorr-fig} and \ref{plots}, however we neglect it in the theoretical expressions below for clarity.

The Lorentzian filter C2 in the idler channel produces a signal photon with the TMF in the shape of a \emph{rising} exponential that terminates at the trigger event (detailed further in the Supplementary Material):
\be
\label{TMF}
\phi(t)= \sqrt\gamma e^{\gamma t /2} \Theta(-t),
\ee
where $\Theta(t)$ is the Heaviside step function and $\gamma=2\pi\times 7$ MHz is the narrowband cavity linewidth. This exponentially rising mode is of particular significance for applications such as high-efficiency excitation of an atom \cite{kur13,kur14} or a resonator \cite{Bad13,Du14} with a single photon. To our knowledge, this is the first demonstration of complete reconstruction of this mode.

\vspace{2 ex}
\noindent\textbf{Results and Discussion}\\

Fig.~\ref{plots}(a) shows the TDM obtained by iterative reconstruction from the experimental data along with the theoretical predictions. The primary eigenvector of the TDM has a corresponding eigenvalue almost 45 times larger than the second largest one indicating a nearly pure temporal mode. The TDM is primarily real and matches well the theoretical prediction.

Next, we demonstrate the reconstruction of a temporal mode with a nonvanishing imaginary component. To this end, we induce a virtual frequency shift by redefining the signal-LO detuning according to $\dom^\prime = \dom + \Delta$ when reconstructing the TDM from Eq.~\eqref{autocorr-theo}. The theoretically expected TMF and TDM then become:
\begin{eqnarray}
\label{TMFshifted}
\phi_\Delta(t) &=& \phi(t) e^{i \Delta t};  \\
\label{TDMshifted}
\rho_\Delta(t,t^\prime) &=&\rho(t,t^\prime) e^{i\Delta (t-t^\prime)}.
\end{eqnarray}
The TDM reconstructed from the experimental data using the effective modulation frequency of $\Delta=2\pi\times 5$ MHz is shown in Fig.~\ref{plots}(b). While the purity of the temporal mode is maintained, the reconstructed density matrix now has a significant imaginary component, demonstrating the ability of our technique to accurately reconstruct states with complex temporal modes.

This example is practically relevant in a situation when one does not know the frequency of the photon precisely. In this case, heterodyne measurements at different LO detunings relative to a given reference (defined by the point $\delta\omega=0$) will provide full information about the mode, including the spectral offset of the photon with respect to that reference.

%Note also that, if only a real part of the TDM reconstructed here were used for homodyne tomography, a quantum efficiency of ??? would be obtained, whereas

Finally, we illustrate the ability of our experimental technique to reconstruct the  TDM in the case of a mixed state. We phase modulate the signal photons at a frequency $\omega_m=2\pi\times 20$ MHz, larger than the spectral width of C2. This is achieved by passing the idler photons through an electro-optic modulator (EOM), with its optical axis oriented along the photon's polarization \cite{Har08}. This leads to a TMF $\phi^{\rm EOM}(t,\theta_m) = \sqrt\gamma e^{\gamma t/2} e^{i\beta \sin(w_m t + \theta_m)}\Theta(-t)$, where $\beta=1.1$ is the modulation index and $\theta_m$ is the phase of the modulating voltage at the time when the idler photon is detected. Because the idler photon detections occur at random times, $\theta_m$ is randomized, leading to the following non-pure TDM:
\begin{eqnarray}\label{TDMEOM}
\rho^{\rm EOM}_{t,t^\prime} &=& \frac 1 {2\pi} \int_{-\pi}^\pi \phi^{\rm EOM}(t,\theta_m)[\phi^{\rm EOM} (t^\prime,\theta_m)]^* d \theta_m\\
&=& \gamma e^{\frac{\gamma(t+t^\prime)}{2}}  \Theta(-t)\Theta(-t^\prime)  J_0 \left[2\beta \sin\left(\frac{\omega_m (t-t^\prime)}{2}\right)\right],\nonumber
\end{eqnarray}
where $J_0$ is the Bessel function of the first kind.

The experimentally reconstructed TDM is shown in Fig.~\ref{plots}(c). The mixed nature of the density matrix is evident from the distribution of eigenvalues, with the ratio of the first and second eigenvalues being only around 2. Due to the modulation phase randomization, the imaginary part of the density matrix is zero. %\emph{In this figure, we also plot the diagonal of the TDM, which is not affected by the modulation as expected from \eeqref{TDMEOM}. %This is the temporal shape that one would obtain by acquiring the photon detection timing statistics.  }
%However, this is just a technical limitation and can be overcome by recording the EOM modulation voltage phases simultaneously with the photocurrent.

%\noindent%
%\begin{minipage}{\linewidth}% to keep image and caption on one page
%\makebox[\linewidth]{%        to center the image
%  \includegraphics[keepaspectratio=true,scale=0.6]{dmplot}}
%\captionof{figure}{{\label{plots}Experimentally obtained TDM (a, c, e) and TMF (b, d, f), with real part (left) and imaginary part (right) for the following cases: without modulation (a, b), with virtual phase modulation in a frequency-shifted frame (c, d), and with EOM modulation (e, f). The insets of the TDM show corresponding theoretical plots. The TMF plotted is the primary eigenvector as reconstructed from experimental data (solid red) and theoretical modeling (dashed blue). The insets in the TMF plots (real) show the distribution of eigenvalues (red bars) obtained experimentally. In the case with EOM modulation (f), the TDM is mixed and the solid blue dots in inset of dominant eigenvector (real) represent the theoretical eigenvalues versus the mode number.}%      only if needed
%\end{minipage}

The observed artifacts in the reconstructed photon modes can be attributed to the finite bandwidth, or non-instantaneous response, of the homodyne detector. This results in the smearing of the acquired autocorrelation matrix. This effect is particularly significant where this matrix has sharp features, such as the trigger event where the photon pulse instantly terminates according to Eq.~\ref{TMF}, as seen in Fig.~\ref{plots}(a,b). The fast modulation of the TMF such as in Fig.~\ref{plots}(c) has a similar effect on the reconstructed mode over its entire duration, resulting in a spurious nonzero imaginary part. The observed artifacts, however, do not significantly degrade the fidelities of the experimentally obtained TDMs with respect to the theoretically expected ones. These fidelities, defined as $F = \tr[\sqrt{\sqrt{\rho_{\rm exp}}\rho_{\rm th}\sqrt{\rho_{\rm exp}}}]$ with the subscripts indicating theory versus experiment, are found to be 0.97, 0.94 and 0.93, respectively, for the three cases of Fig.~\ref{plots}.

Using the absolute value of the TMF obtained for the primary mode of the unmodulated case [Fig.~\ref{plots}(a)], we reconstruct the quantum state of light in that mode in the Fock basis akin to Ref.~\cite{And12PRL}, obtaining the single-photon efficiency of $\rho_{11}=52.9$\%. The corresponding Wigner function, exhibiting negative values at the phase-space origin, is plotted in Fig.~\ref{exptsetup}(c) along with the acquired quadrature distribution and the reconstructed density matrix. %This shows a relative improvement (around 8$\%$) over our previous work \cite{And12PRL} where the temporal mode was simply estimated theoretically, emphasizing the importance of knowing the temporal mode precisely for realizing high-efficiency quantum measurement and interactions.

\vspace{2 ex}
\noindent\textbf{Conclusion}\\
We have developed and experimentally demonstrated polychromatic optical heterodyne tomography, a robust method for complete experimental determination of the temporal properties of a single photon directly from the time-resolved photocurrent statistics of a balanced homodyne measurement. The method enables the extraction of a temporal mode which in general may be complex and can have multiple frequency components. Accurate detection of the temporal mode is key for the proper mode matching required by many quantum communication protocols. %Subsequent tomographic reconstruction of the quantum optical state in the evaluated mode permits one to evaluate the quality of single-photon sources.

Our method permits straightforward extension to states other than the single-photon Fock state akin to Ref.~\cite{Mor13} provided that the state in question occupies a well-defined spatiotemporal mode. On the other hand, the single-photon state is special in that it can be directly associated with the photon annihilation operator of a certain optical mode or a mixture thereof. The problem of defining the optical mode(s) for a general quantum optical state is a subject of a separate study.

Although the technique described in this work requires the frequency spectrum of the photon's temporal mode to be sufficiently narrowband so its temporal structure can be resolved by the homodyne  detector, one can envision ways to lift this restriction. For example, if the photon is produced in an ultrashort pulsed mode, one can extend it in time using a dispersive element such as an optical fiber, and perform time-resolved homodyne detection using a matched chirped local oscillator. The time-domain correlations of the homodyne photocurrent will then correspond to quantum coherences between components of the photon spectrum.

\vspace{2 ex}
\noindent\textbf{Acknowledgements}\\
ZQ and ASP contributed equally to this work. We thank Erhan Saglamyurek and Wolfgang Tittel for lending us the EOM. The project is supported by NSERC and CIFAR. AL is a CIFAR Fellow. ZQ is supported by the China Scholarship Council.

\nocite{*}

\vspace{2 ex}
%\noindent\textbf{References}
%

\makeatletter
\renewcommand{\thefigure}{S\@arabic\c@figure}
\makeatother
\setcounter{figure}{0}
%\newpage
\section{Appendix}
\subsection{Derivation of temporal wavefunction}

The theoretical TMF for the single photon can be obtained as follows. The biphoton generated by the four-wave mixing process with an infinite gain bandwidth can be represented as
\be
\label{TMSS}
\ket \Psi = \int \ket{1_{\omega_s}, 1_{\omega_i}} \delta(\omega_s + \omega_i - 2\omega_p) d\omega_s d\omega_i
\ee
where indices $s$, $i$ and $p$ correspond to signal, idler and pump respectively. The amplitude transmission function of the cavity in the neighborhood of its resonance can be expressed as a function of the detuning of the beam from the cavity resonance frequency $\delta_i = \omega_c-\omega_i$ as $T(\delta_i)= \sqrt{2/\pi\gamma}(1-2i\delta_i/\gamma)^{-1}$, where $\gamma$ is the linewidth of the cavity. Subjecting the idler channel to transmission through this cavity, state (\ref{TMSS}) transforms into:
%a subsequent single photon detection will project the signal onto the state $\ket {\tilde{\phi}_s} = \Braket{t_i|\psi^\prime}$ where $\psi^\prime$ is given by
\be
\label{TMSSaftercav}
\ket{\Psi^\prime} = \sqrt{\frac 2{\pi\gamma}}\int \frac{1}{1-2i\frac{\delta_i}{\gamma}} \delta(\omega_s + \omega_i - 2\omega_p) \ket{1_{\omega_s}, 1_{\omega_i}}  d\omega_s d\omega_i.
\ee
Subsequent idler photon detection at time $t_i$ will project the signal onto the state $\ket {\tilde{\phi}_s} = \langle 1_{t_i}|\Psi^\prime\rangle$ where $\ket{1_{t_i}} = \int \ket{1_{\omega_i}} e^{i \delta_i t_i} d\omega_i$. Assuming that the detection event happens at $t_i=0$, the resultant state in the signal channel in the frequency domain can be expressed as
\be
\label{projection}
\ket{1_{\phi_s}} = \langle1_{t_i=0}|\Psi^\prime\rangle = \sqrt{\frac 2{\pi\gamma}}  \int \frac{1}{1 + 2i\frac{\delta_s}{\gamma}} \ket{1_{\omega_s}} d\omega_s,
\ee
where $\delta_s= 2\omega_p - \omega_c - \omega_s$ is the detuning of the signal frequency from the central frequency determined by the cavity. Performing a Fourier transform on Eq.~(\ref{projection}), we find the temporal mode of the signal photon:
\begin{equation}
\label{TMF-sl}
\ket{1_{\phi_s}} = \sqrt\gamma \int e^{\gamma t_s /2} \Theta(-t_s) \ket{1_{t_s}} dt_s
\end{equation}
where $\Theta(\cdot)$  is the step function. %This rising exponential mode is similar to that of a cavity enhanced photon studied in Ref.~\cite{Bad13}.

\subsection{Ambiguity of reconstruction with a single LO frequency}
\begin{figure}[htbp]
\includegraphics[width=\columnwidth]{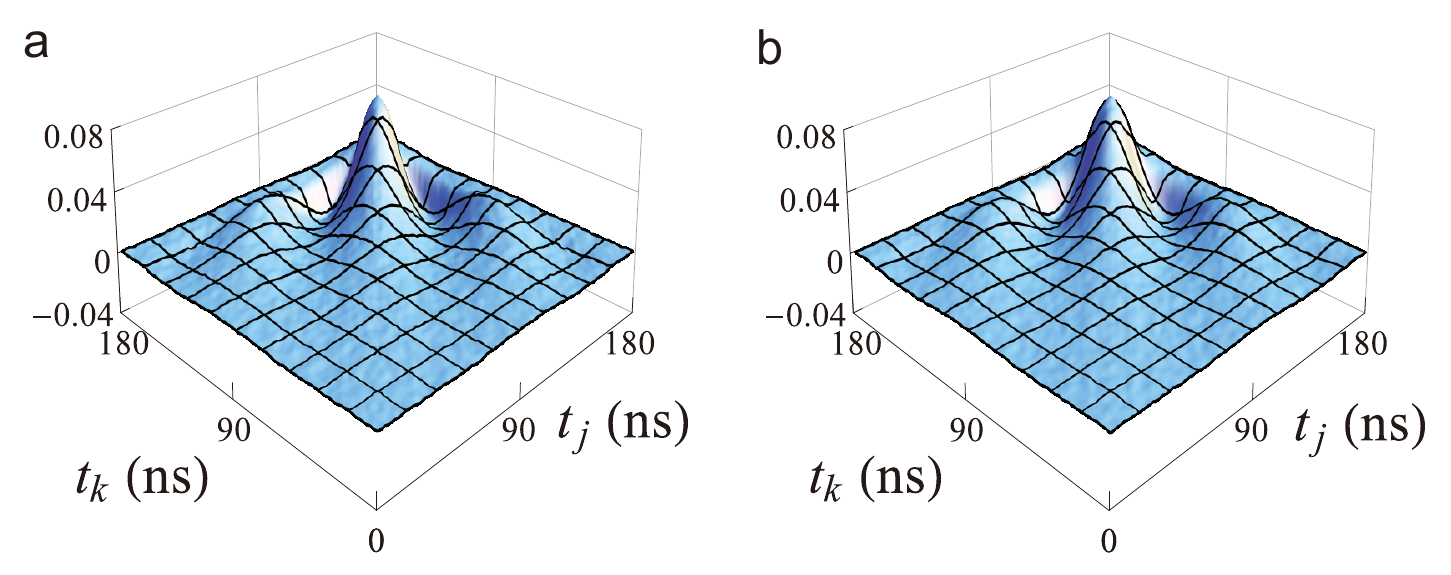}
\caption{\label{PlusMinusDelta}Experimentally observed reduced autocorrelation matrices $\re\rho_\Delta (t,t^\prime)$ for $\Delta=10.8$ MHz (a) and $\Delta=-10.8$ MHz (b). These matrices are identical, although the temporal modes are different.}
\end{figure}
If only a single LO frequency with $\delta\omega=0$ were used, as in Refs.~\cite{And12PRLsup,Mor13sup}, mode reconstruction would only yield a real part of the TDM. To present direct evidence of incompleteness of such reconstruction, we specialized to the situation of Fig.~3(b) (virtual phase modulation) and acquired the homodyne photocurrent autocorrelation for the photon detuned from the local oscillator by $\Delta=\pm 10.8$ MHz (Fig.~\ref{PlusMinusDelta}). The TDMs corresponding to these situations are complex conjugate of each other: $\rho_\Delta (t,t^\prime)= \gamma e^{\gamma (t+t') /2} \Theta(-t)
\Theta(-t')e^{\pm i \Delta (t-t^\prime)}$. However, the reduced autocorrelation matrix $\re\rho_\Delta (t,t^\prime)$ is the same  in both these cases or for their arbitrary statistical mixture.

%\subsection{Significance of the imaginary component of the TDM}
%Here we present an explicit example of how a single LO frequency autocorrelation matrix is insufficient for the unambiguous mode reconstruction. To that effect, acquire auto

\subsection{Calculating reduced autocorrelation matrices}
\begin{figure}[htbp]
\includegraphics[width=\columnwidth]{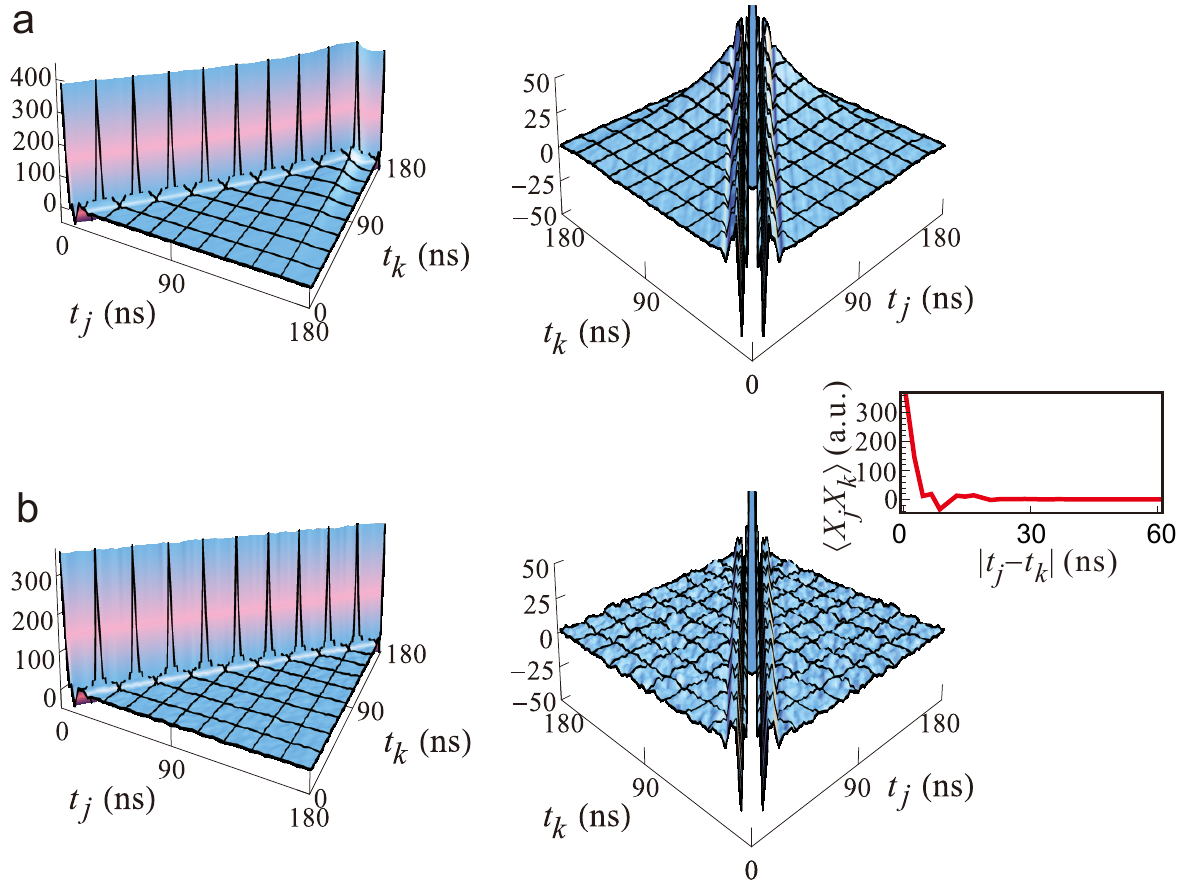}
\caption{\label{FullAuto}Experimentally observed autocorrelation matrices of the homodyne detector photocurrent for (a) the single-photon signal with zero LO detuning, no modulation and (b) vacuum signal. Two different viewpoints and plot ranges are used for better visibility. The reduced autocorrelation matrix [Fig.~2(b) of the main text] is obtained by subtracting (b) from (a). The inset shows the mean homodyne detector photocurrent autocorrelation $\langle X_j X_k \rangle$ for the vacuum case as a function of $|t_j-t_k|$. The vertical axis scale is arbitrary.}
\end{figure}

The experimentally obtained full autocorrelation matrices are shown in Fig.~\ref{FullAuto} for the signal being in the single-photon (a) and vacuum (b) states. These matrices are described by Eqs.~(5) and (6) in the main text. The matrix in (a) contains  both terms of Eq.~(5), while that of (b) only the first term. One can see that the first term is dominant, while the ``payload" second term emerges only as a small feature in Fig.~\ref{FullAuto}(a). This is to be expected: according to Eq.~(5), the magnitude of the first term is $1/2$ whereas that of the second term is determined by the typical TDM element, which is the inverse duration of the photon's temporal mode measured in digitizer time bins (because the trace of the density matrix must be one). As seen from Fig.~3 in the main text, the largest elements of the normalized density matrix do not exceed 0.06--0.07. Note that using excessive temporal resolution of the digitizer is therefore not advisable, as it will reduce the ratio between that density matrix and the vacuum term.

Additionally, the experimental autocorrelation matrices are affected by the statistical uncertainties. For each waveform sample and a specific pair of time bins, the variance of the autocorrelation is evaluated for the vacuum case as
$$\langle\Delta(X_jX_k)^2\rangle=\langle X_j^2\rangle\langle X_k^2\rangle=\frac 1 4.$$
For $N$ waveforms acquired, the standard deviation of the mean autocorrelation value is therefore $1/(2\sqrt N)$. For the single-photon case, the statistical noise is of the same magnitude. For reliable reconstruction of the mode, this noise must be much less than the typical density matrix element. We address this issue by acquiring large quantities of data samples: $N=2\times 10^6$ waveforms for each LO frequency. This is much more than the number of samples collected in a typical quantum state tomography experiment.

In practice, both terms of Eq.~(5), as well as the statistical fluctuations, are smeared by the non-instantaneous response of the detector (Fig.~\ref{FullAuto}). Fortunately, this smearing primarily affects the undesired first term and the fluctuations because of their singular nature.

\begin{figure*}[htbp]
\includegraphics[width=\textwidth]{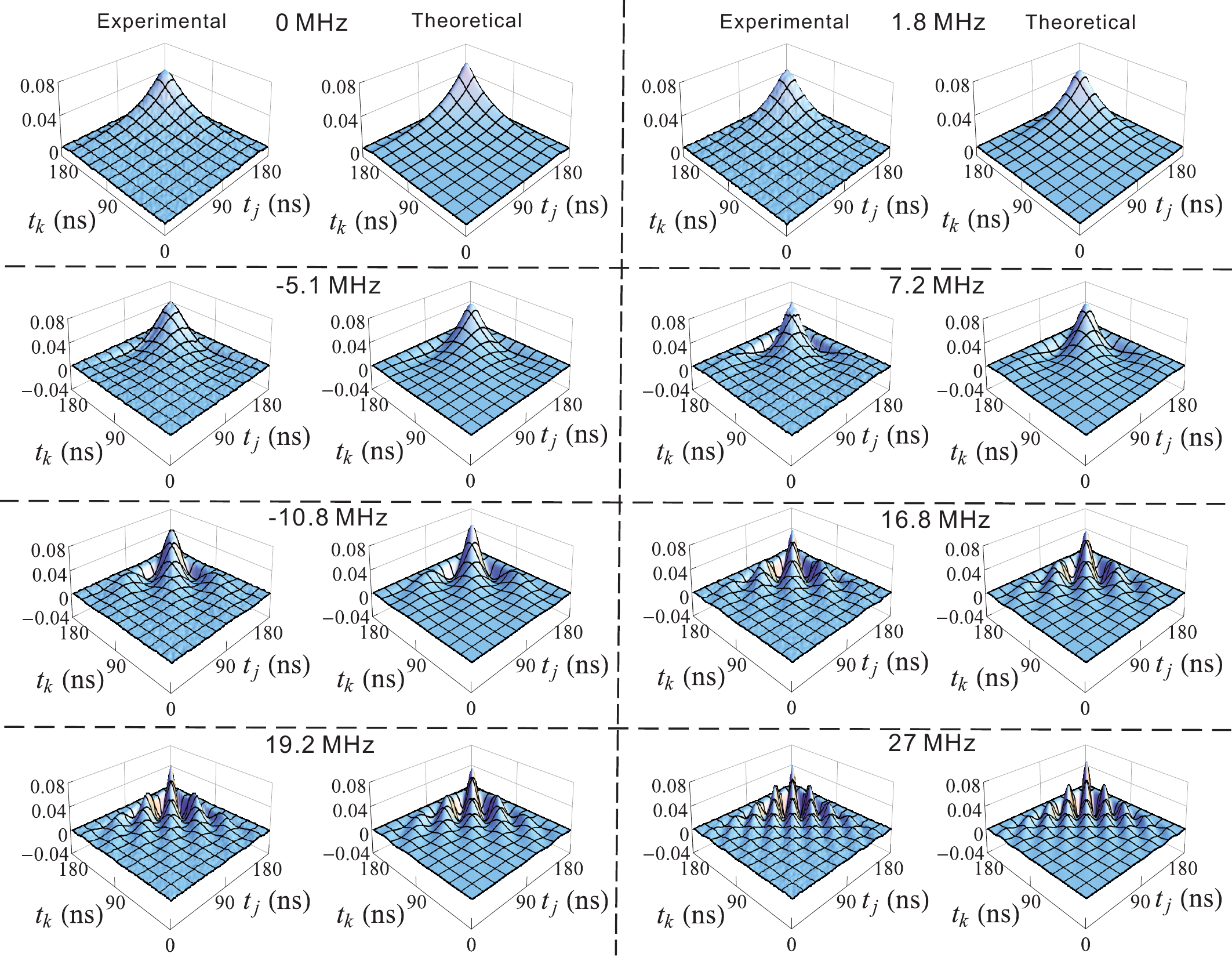}
\caption{\label{ReduAuto}Theoretical (right) and experimental (left) reduced autocorrelation matrices, for eight different LO detunings, corresponding to the measurement setting without modulation. The trigger photon arrives at $t=155$ ns.}
\end{figure*}

\subsection{A full set of reduced autocorrelation matrices}
Eight normalized reduced autocorrelation matrices for the case without modulation are presented in Fig.~\ref{ReduAuto}, and the theoretically calculated reduced autocorrelation matrices are also shown for comparison.

\nocite{*}
%\noindent\textbf{References}
%\vspace{5 ex}
%

\end{document}